\documentclass[sigconf]{acmart}

\usepackage{booktabs} 
\usepackage{arydshln} 
\usepackage{multirow}
\usepackage[ruled]{algorithm2e} 
\usepackage{color}

\SetAlFnt{\small}
\SetAlCapFnt{\small}
\SetAlCapNameFnt{\small}
\SetAlCapHSkip{0pt}

\setcopyright{none}
\renewcommand\footnotetextcopyrightpermission[1]{}
\authorsaddresses{}
\usepackage[T1]{fontenc}
\DeclareRobustCommand{\augiefamily}{%
  \fontfamily{augie}\fontseries{m}\fontshape{n}\selectfont}
\DeclareTextFontCommand{\textaugie}{\augiefamily}

\DeclareCaptionFormat{cont}{#1 (cont.)#2#3\par}

\definecolor{url_color}{RGB}{42, 83, 163}

\makeatletter
\DeclareRobustCommand\onedot{\futurelet\@let@token\@onedot}
\def\@onedot{\ifx\@let@token.\else.\null\fi\xspace}

\def\eg{e.g\onedot} 

\def\ie{i.e\onedot}

\def\etc{etc\onedot}

\def\etal{et al\onedot}
\makeatother

\setcopyright{acmlicensed}
\copyrightyear{2024}
\acmYear{2024}
\acmConference[SIGGRAPH Conference Papers '24]{Special Interest Group on Computer Graphics and Interactive Techniques Conference Conference Papers '24}{July 27-August 1, 2024}{Denver, CO, USA}
\acmBooktitle{Special Interest Group on Computer Graphics and Interactive Techniques Conference Conference Papers '24 (SIGGRAPH Conference Papers '24), July 27-August 1, 2024, Denver, CO, USA}
\acmDOI{10.1145/3641519.3657425}
\acmISBN{979-8-4007-0525-0/24/07}

\begin{document}
% Title portion
\title{
Coin3D: Controllable and Interactive 3D Assets Generation with Proxy-Guided Conditioning
}

\author{
Wenqi Dong$^{1,2}$\footnotemark[1]\footnotemark[2] \quad Bangbang Yang$^2$\footnotemark[1] \quad Lin Ma$^2$ \quad Xiao Liu$^2$ \quad Liyuan Cui$^1$ \\
Hujun Bao$^1$ \quad Yuewen Ma$^2$ \quad Zhaopeng Cui$^{1}$\footnotemark[3]}
\affiliation{
\institution{$^1$Zhejiang University \quad $^2$ByteDance}
\country{}
}

\renewcommand{\shortauthors}{Dong et al.}

\begin{teaserfigure}
  \centering
    \includegraphics[width=1.0\linewidth]{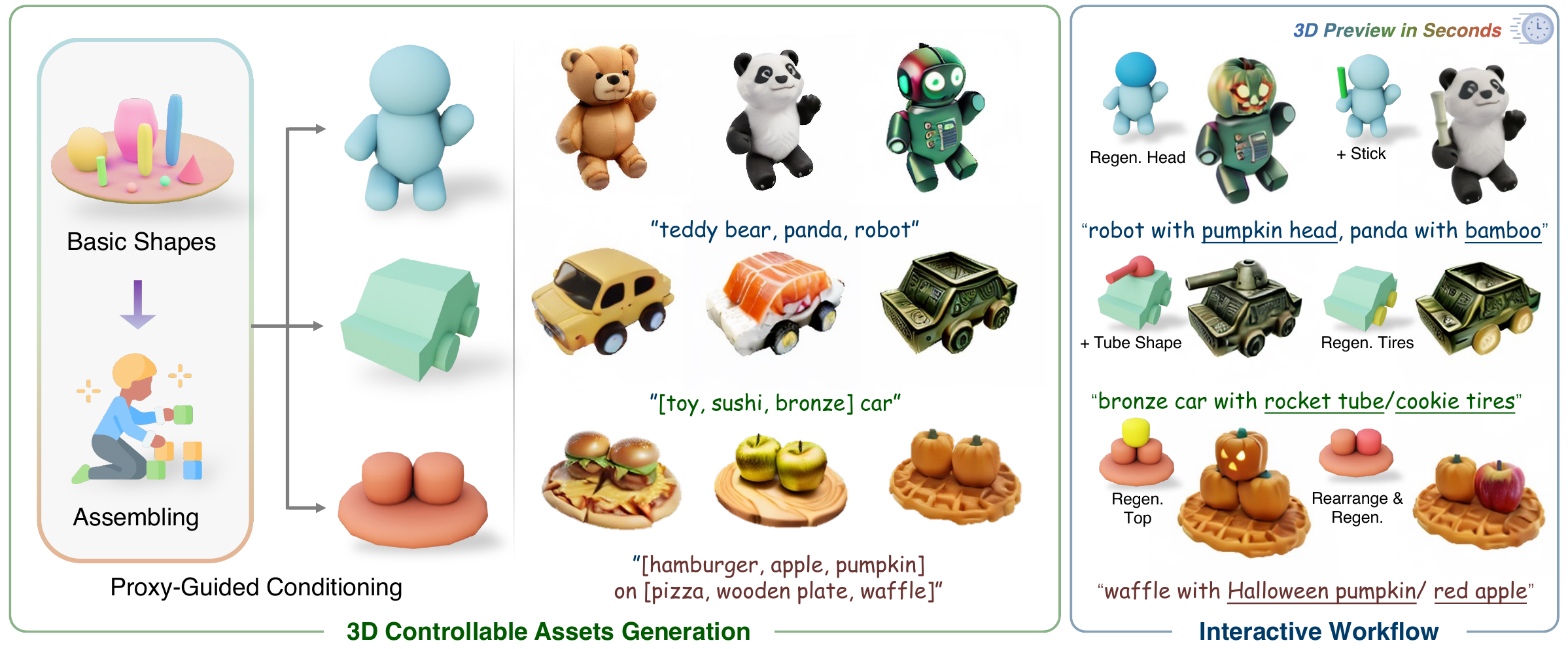}
    \caption{
    Coin3D allows users to add 3D-aware control to the object generation using coarse proxies assembled from basic shapes, enabling an interactive generation workflow with fine-grained part editing and responsive 3D previewing.
    }
    \label{fig:teaser}
\end{teaserfigure}

\begin{abstract}

As humans, we aspire to create media content that is both freely willed and readily controlled.
Thanks to the prominent development of generative techniques, we now can easily utilize 2D diffusion methods to synthesize images controlled by raw sketch or designated human poses, and even progressively edit/regenerate local regions with masked inpainting. 
However, similar workflows in 3D modeling tasks are still unavailable due to the lack of controllability and efficiency in 3D generation.
In this paper, we present a novel controllable and interactive 3D assets modeling framework, named Coin3D.
Coin3D allows users to control the 3D generation using a coarse geometry proxy assembled from basic shapes, and introduces an interactive generation workflow to support seamless local part editing while delivering responsive 3D object previewing within a few seconds.
To this end, we develop several techniques, including the 3D adapter that applies volumetric coarse shape control to the diffusion model, proxy-bounded editing strategy for precise part editing, progressive volume cache to support responsive preview, and volume-SDS to ensure consistent mesh reconstruction.
Extensive experiments of interactive generation and editing on diverse shape proxies demonstrate that our method achieves superior controllability and flexibility in the 3D assets generation task.
Code and data are available on the project webpage:
\urlstyle{tt}
\textcolor{url_color}{\url{https://zju3dv.github.io/coin3d/}}.
\end{abstract}




\maketitle

\renewcommand{\thefootnote}{\fnsymbol{footnote}}
\footnotetext[1]{Wenqi Dong and Bangbang Yang contributed equally to this work.}
\footnotetext[2] {Wenqi Dong conducted this work during his internship at PICO, ByteDance.}
\footnotetext[3]{Corresponding authors.}

\section{Introduction}

As a child, we are born with the instinct to create things with our imagination, \ie, building houses or vehicles using different Lego bricks, or doodling pictures with pencils~\cite{nath2014construction}.
Yet only a few people learn drawing or modeling skills, which eventually develop the ability to create qualified artworks.
Fortunately, the rapid development of generative techniques grants everyone a chance to create fantasy content, \ie, using LLM for automatic manuscripting~\cite{wei2022chain,gpt} or 2D diffusion methods for text-to-image/video generation~\cite{stable_diffusion,lugmayr2022repaint,guo2023animatediff}.
To enable the controllability of the generative model, recent advances in 2D diffusion (such as ControlNet~\cite{controlnet}, T2I-Adapter~\cite{mou2023t2i}, SDEdit~\cite{meng2021sdedit}, and \etc) allow users to take depth, sketches or human poses to control the generation process, enables iteratively editing designated region with inpainting and re-generating mechanism.
However, in the field of 3D asset generation~\cite{dreamfusion}, existing 3D generative methods still lack controls for artistic creation.
First, they are usually conditioned with text prompts~\cite{dreamfusion} or perspective images~\cite{zero123,magic123,a12345,wonder3d,syncdreamer}, which is not sufficient to express 3D objects accurately.
Second, when performing high-level tasks such as generative editing~\cite{li2023focaldreamer,cheng2023progressive3d,haque2023instruct,kamata2023instruct} or inpainting~\cite{zhou2023repaint}, existing approaches usually require a significant amount of time for reconstruction before previewing the editing operation.

Given this observation, we believe that a controllable and user-friendly 3D assets generation framework should have these properties.
\textbf{(1) 3D-aware controllable}: similar to a child stacking Lego bricks and picturing the vivid appearance in its mind, a controllable 3D generation can be started by assembling basic shapes (\eg, cuboids, spheres, or cylinders), which serves as coarse shape guidance for the detailed generation.
Therefore, it reduces the difficulty of 3D modeling for common users and also provides sufficient control over the generation.
\textbf{(2) flexible}: the framework should allow users to interactively composite or adjust local regions in a 3D-aware manner, ideally as easy as image inpainting tools~\cite{meng2021sdedit}.
\textbf{(3) responsive}: once the user's editing is temporarily finished, the framework should instantly deliver preview images of the generated object from the desired viewpoints, rather than waiting for a long reconstruction period.
In this paper, we propose a novel \textbf{CO}ntrollable and \textbf{IN}teractive 3D assets generation framework, named Coin3D.
Instead of using text prompts or images as conditions,
Coin3D allows users to add 3D-aware conditions into a typical multiview diffusion process in the 3D generation task, \ie, using a coarse 3D proxy assembled from basic shapes to guide the object generation, as illustrated in Fig.~\ref{fig:teaser}.
Based on proxy-guided conditioning, Coin3D introduces a novel generative and interactive 3D modeling workflow.
Specifically, users can depict the desired object by typing in text prompts and assembling basic shapes with their familiar modeling software (such as Tinkercad, Blender, and SketchUp).
Then, Coin3D would construct the on-the-fly feature volume in a few seconds, which enables the preview of the result from arbitrary viewpoints or even progressively adjust/regenerate the designated local part of the object.
For example, we can generate a bronze car by assembling basic shapes and incrementally adding tubes or changing tires as shown in Fig.~\ref{fig:teaser}.
However, even though adding 3D-aware conditions is technically plausible, there are still some challenges to an interactive 3D modeling workflow, which will be addressed in this work:

\paragraph{Coarse Shape Guidance.}
Since we only use simple basic shapes (\eg, stacked spheres or cuboids) instead of intricate CAD models for 3D guidance, the proxy-guided conditioning should allow some freedom during the generation rather than being strict to the given basic shapes, \eg, growing animal ears from the sphere head as shown in Fig.~\ref{fig:teaser}.
To achieve this goal, we design a novel 3D adapter to process the 3D control, where the 3D proxies (basic shapes) are first voxelized and extracted to 3D features, and then integrated into the spatial features of a multiview generation pipeline~\cite{syncdreamer}.
In this way, users can manipulate the control strength by changing the plug-in weights, enabling controlling the generated object more or less close to the given proxy.

\paragraph{Interactive Modeling.}
An interactive and productive 3D generation workflow should support progressive modeling operations and responsive preview, \ie seamlessly adding/adjusting shape primitives or precisely regenerating local parts without touching others, while all the operated results should be previewed as quickly as possible without time-consuming reconstruction.
To fulfill the demands, we first develop a novel proxy-bounded editing strategy, which ensures precise bounded control and natural style blending when modifying part of the object, and then utilize a progressive volume caching mechanism by memorizing stepwise 3D features to enable responsive preview. 

\paragraph{Consistent Reconstruction.}
To facilitate the standard CG workflow, one might need to export the generated assets into textured mesh with reconstruction.
However, even with 3D-aware conditioning, there might still be poor reconstructions when na\"ively reconstructing objects using synthesized multiview images due to the limited viewpoints (see Sec.~\ref{ssec:exp_ablation}).
To tackle this issue, we propose to leverage the proxy-guided feature volume during reconstruction with a novel volume-SDS loss.
This strategy effectively exploits the controlled 3D context during the score distillation sampling~\cite{dreamfusion} and faithfully improves the reconstruction quality.

Our contributions can be summarized as follows.
\textbf{1)}
We propose a novel controllable and interactive 3D assets generation framework, named Coin3D.
Our method designs a 3D-aware adapter to take simple 3D shape proxies as guidance to control the object generation, which supports interactive generation operations such as altering prompts, adjusting shapes, or fine-grained local part regeneration. 
\textbf{2)}
To ensure an interactive and consistent experience of generative 3D modeling, we develop several techniques, including proxy-bounded editing for precise and seamless part editing, progressive volume cache to support responsive preview from arbitrary views, and a conditioned volume-SDS to improve the mesh reconstruction quality.
\textbf{3)}
Extensive experiments of interactive generation with various shape proxies and the interactive workflow deployed on the 3D modeling software (\eg, Blender) demonstrate the controllability and productivity of our method on generative 3D modeling.

\section{Related Works}

\begin{figure*}[t!]
    \centering
    \includegraphics[width=1.0\linewidth, trim={0 0 0 0}, clip]{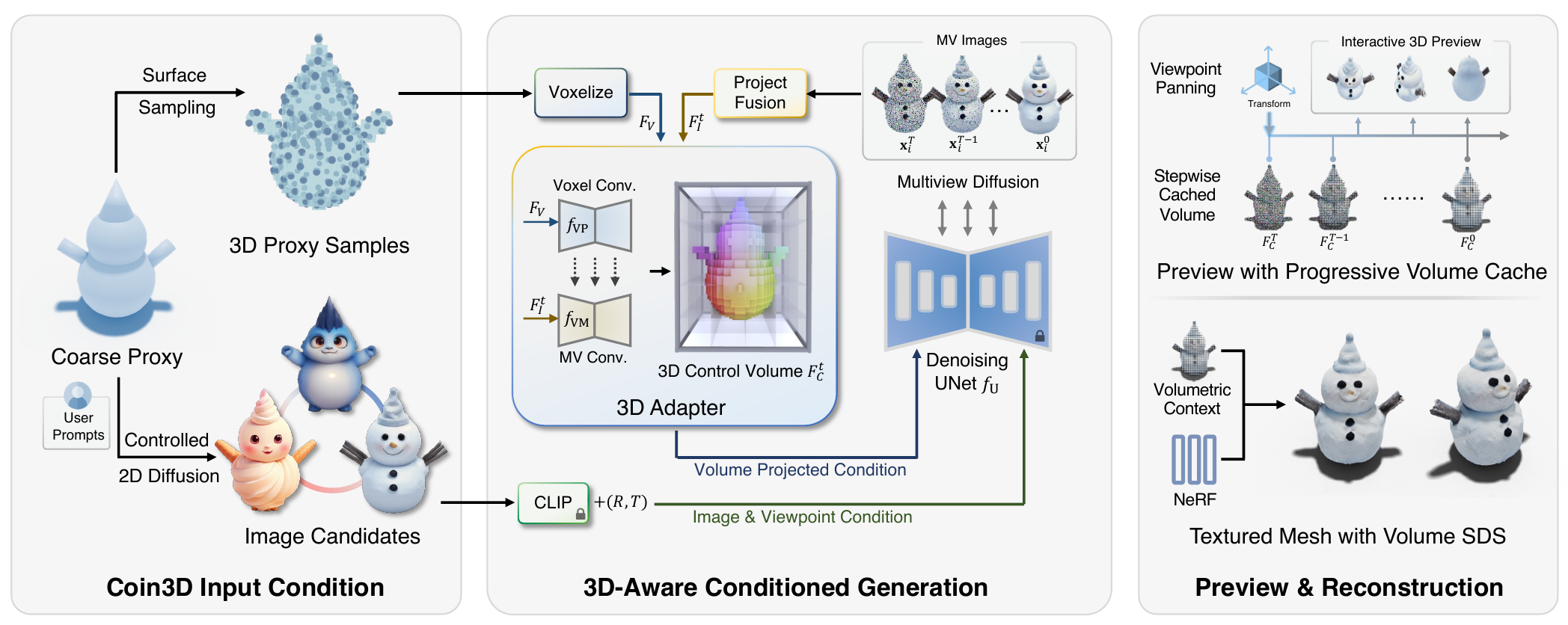}
    \caption{
    \textbf{Overview.}
    Given a coarse shape proxy and user prompts that describe the identity, our method first constructs 2D image candidates from the proxy's silhouette and 3D proxy samples as input conditions.
    Then, we employ a 3D adapter to integrate 3D-aware control to the diffusion's denoising process with a 3D control volume $F_C$, yielding multiview images of the object.
    By fully leveraging $F_C$, we realize accelerated 3D previewing with volume cache and also improve mesh reconstruction quality.
    }
    \label{fig:framework}
\end{figure*}

\subsection{3D Object Generation}
3D object generation is a popular task in computer vision and graphics.
Early works~\cite{dubrovina2019composite, achlioptas2018learning, kluger2021cuboids} mainly focus on natively generating 3D representations from models, such as polygon meshes~\cite{gao2022get3d,groueix2018papier,kanazawa2018learning,nash2020polygen,wang2018pixel2mesh}, pointclouds~\cite{achlioptas2018learning,fan2017point,nichol2022point,yu2023pushing}, parametric models~\cite{hong2022avatarclip,text2human}, voxels~\cite{choy20163d,sanghi2022clip,wu2017marrnet,xie2019pix2vox}, or implicit fields~\cite{cheng2023sdfusion,jun2023shap,li2023diffusion,mescheder2019occupancy,park2019deepsdf, chan2021pi, gu2021stylenerf, chan2022efficient, skorokhodov2022epigraf}, which learn from specific CAD database~\cite{shapenet} and are often bounded by specific categories (\eg, chairs, cars, and \etc) due to the limited network capacity and data diversity.
Recently, with the rapid evolution in large-scale generative models, especially the great success in 2D diffusion models~\cite{imagen, dalle2, stable_diffusion}, methods like DreamFusion~\cite{dreamfusion}, SJC~\cite{wang2023score} and their follow-up works~\cite{fantasia3d,magic3d,melas2023realfusion,raj2023dreambooth3d,seo2023let,tang2023dreamgaussian,tang2023make,xu2023dream3d} attempt to distill 2D gradient priors from the denoising process using score distillation sampling loss (SDS loss) or its variants, which guide the per-shape neural reconstruction following users' text prompts.
While being generic to unlimited categories and diverse composited results with prompt engineering, these lines of work often suffer from unstable convergence due to the noisy and inconsistent gradient signal, which often leads to incomplete results or ``multi-face Janus problem''~\cite{fantasia3d}.
Subsequently, Zero123~\cite{zero123} analyzes the viewpoint bias problem of the generic 2D latent diffusion model (LDM), and proposes to train an object-specific LDM with relative viewpoint as a condition using Objaverse dataset~\cite{objaverse}, which shows promising results in the image-to-3D tasks, and has been widely adopted in the follow-up 3D generation works~\cite{magic123,a12345}.
While being fine-tuned on multiview images, Zero123 still suffers from the cross-view inconsistency issue as its resulting images cannot satisfy the requirements for reconstruction.
Hence, later works such as MVDream~\cite{mvdream}, SyncDreamer~\cite{syncdreamer}, Zero123++~\cite{zero123++} and  Wonder3D~\cite{wonder3d} propose to enhance multiview image generation, which either trains with stacked views~\cite{mvdream,wonder3d,zero123++}
or builds synchronized volumes online to condition the diffusion process~\cite{syncdreamer},
and usually enables to produce highly consistent images or yields 3D reconstructions in few seconds.
Very recently, LRM~\cite{lrm} and its variant methods~\cite{pf_lrm,dmv3d} propose to train an end-to-end transformer-based model, which directly produces neural reconstruction given one or few perspective images.
Nevertheless, existing 3D object generation methods primarily focus on using text prompts (text-to-3D) or images (image-to-3D) as the input, which cannot accurately convey exact 3D shapes or precisely control the generation in a 3D manner.
By contrast, our method first adds 3D-aware control to the multiview diffusion process without compromising generation speed, which realizes interactive generation workflow with 3D proxy as conditions.

\subsection{Controllable and Interactive Generation}

Adding precise control to the generative methods is crucial for productive content creation~\cite{epstein2022blobgan,bao2023sine,object_nerf,nr_in_a_room,neumesh,dreamspace}.
Previous generative works~\cite{melnik2024face, chen2022sem2nerf, deng20233d, hao2021gancraft,bao2024geneavatar} mainly learn a latent mapping of the attributes to add control to the generation, but are limited to specific categories (\eg, human faces or nature landscape).
Recent progress in 2D diffusion models, such as ControlNet~\cite{controlnet} and T2I-Adapter~\cite{mou2023t2i}, enables various 2D image hints (\eg, depth, normal, softedge, human poses, color grids, and \etc), to interactively control the denoising process of the image generation.
However, similar controllable capabilities~\cite{pandey2023diffusion, bhat2023loosecontrol, cohen2023set} in 3D generation are far from applicable.
For generative 3D editing, recent works~\cite{li2023focaldreamer,cheng2023progressive3d} propose to constrain the text-driven 3D generation at the desired region, but cannot support controlling the exact geometry shape.
For Controllable 3D generation, the most related works to our methods are Latent-NeRF~\cite{latentnerf} and Fantasia3D~\cite{fantasia3d}.
However, these two works cannot ensure steady convergence and the generated results are usually far from the given 3D shape (see Sec.~\ref{ssec:compare_3d_control}), as they na\"ively add control to the 3D representation regardless of altering the supervision of 2D priors (\ie, SDS loss).
Other works such as Control3D~\cite{chen2023control3d} only add control from 2D sketches/silhouettes instead of 3D space.
Moreover, all these methods require a long time of reconstruction (\eg, from dozens of minutes to hours) to inspect the effect of editing or controlling, which cannot fulfill the demand for interactive modeling.
On the contrary, our method directly integrates the 3D-aware control into the diffusion process, which not only ensures faithful and adjustable control over the 3D generation but also allows to interactively preview the generated/edited 3D object in a few seconds.

\section{Method}

We introduce Coin3D, a novel Controllable 3D assets generation framework, which adds 3D-aware control to the multiview diffusion process in object generation tasks, enabling an interactive modeling workflow for fine-grained customizable object generation.
An overview of Coin3D is shown in Fig.~\ref{fig:framework}. 
Instead of using conventional text prompts or a perspective image as a condition, our framework employs a coarse geometry proxy made from basic shapes (\eg, a snowman composed of two spheres, two sticks, and one cone), complemented by user prompts that describe the object's identity.
Then, the diffusion-based generation will be conditioned on both a voxelized 3D proxy and 2D image candidates generated by controlled 2D diffusion with the proxy's silhouette (\eg, images with different appearances in the left bottom of Fig.~\ref{fig:framework}).
During the 3D-aware conditioned generation, we use a novel 3D adapter module that seamlessly integrates proxy-guided controls with adjustable strength into the diffusion pipeline (Sec~\ref{ssec:method_3d_cond}).
To deliver an interactive generation workflow with fine-grained 3D-aware part editing and the responsive previewing ability, we also introduce proxy-bounded editing to precisely control the volume update, and employ an efficient volume cache mechanism to accelerate the image previewing at arbitrary viewpoints (Sec~\ref{ssec:method_interact}).
Furthermore, we propose a volume-conditioned reconstruction strategy, which effectively leverages the 3D context from feature volume to improve the reconstruction quality (Sec~\ref{ssec:method_recon}).

\subsection{Proxy-Guided 3D Conditioning for Diffusion}
\label{ssec:method_3d_cond}

\paragraph{3D proxy as initial condition.}
As illustrated in Fig.~\ref{fig:framework}, our method uses a 3D coarse shape proxy assembled from basic elements (\eg, cubes, cylinders, cones, spheres, \etc) and user prompts to condition the multiview diffusion process.
More specifically, given the coarse shape $P$ and prompts $y$, we want to predict $N_v$ consistent images $\{\textbf{x}_i|i= 1,2,\ldots, N_v\}$ under the camera poses $\{\textbf{c}_i|i= 1,2,\ldots, N_v\}$ using a multiview diffusion-based generator $f$ as the following:%, which can be modeled as:
\begin{equation}
    \textbf{x}_{(i:N_v)} = f(P,y,\textbf{c}_{(i:N_v)}).
\end{equation}
Note that, unlike regular 2D diffusion, multiview diffusion synchronously performs denoising iterations on all the preset views, which integrates cross-view correlations with view-dependent self-attention~\cite{wonder3d,zero123++} or spatial volume~\cite{syncdreamer,a12345}.   
To simplify the preparation of proxy shapes, our method allows the user to realize the input by simply scaling and assembling basic shapes in 3D modeling software (\eg, Tinkercad, SketchUp, or Blender) without relying on complex modeling skills.
Hence, to adapt the coarse proxy inputs (\ie, a coarse polygon mesh) for 3D generation tasks, we develop a two-pathway condition preprocess for the proxy.
First, we sample $N_S$ surface points $\mathcal{P}=\{\textbf{p}_i|i=1,2,\ldots,N_{S}\}$ on the proxy mesh, which will be used for the 3D-aware control for the generation pipeline.
Second, we use the proxy's rendered silhouette and users' prompts as a condition, 
and generate multiple 2D image candidates for interactive appearance selection ~\cite{stable_diffusion,controlnet}.

\paragraph{3D-aware control with 3D adapter.}
We introduce the 3D adapter to add 3D-aware control from coarse proxy samples to the multiview diffusion pipeline, which yields multiview images of the object following the given proxy shape.
To achieve the lossless 3D control of the diffusion model, inspired by volumetric multiview diffusion works~\cite{syncdreamer,a12345++}, 
we construct a 3D control volume to add the 3D-aware context into the diffusion pipeline, where the volume is a voxel feature grid containing $v^3$ vertices.
As shown in the middle part of Fig.~\ref{fig:teaser}, the 3D adapter receives two inputs from proxy feature volume $F_V$ and the multiview image fused volume $F_I^t$.
Specifically, $F_V$ is constructed by first voxelizing the proxy samples $\mathcal{P}$ to fill in the zero-initialized occupancy grid, where each grid will be assigned to $1$ if containing any point.
$F_I^t$ is the multiview feature volume constructed by unprojecting and fusing multiview images $\textbf{x}_{(i:N)}^t$ produced by the denoising UNet $f_{\text{U}}$ at timestamp $t$.
$F_I^t$ is the multiview feature volume, which is constructed by first projecting vertices of $V$ onto the multi-view images $\textbf{x}_{(i:N)}^t$ to obtain interpolated image-plane features, and then fusing them with a 3D CNN module.
$\textbf{x}_{(i:N)}^t$ are produced from the denoising process of UNet $f_{\text{U}}$ at timestamp $t$.
Then, for each denoising step $t$ in the 3D adapter, we first perform 3D convolution (with 3D UNet $f_{\text{VP}}$) on the volume $F_V$, and hierarchically add the intermediate layer outputs to the 3D convolution (with 3D UNet $f_{\text{VM}}$) of multiview feature volume $F_I^t$, which yields the final 3D control volume $F^t_C$.
Then, during the multiview denoising of the 2D diffusion model, we first project the $F^t_C$ to align the corresponding view to obtain 2D feature map $F_{p}^t$~\cite{mvsnet}, and then integrate the $F_{p}^t$ (with depthwise attention) along with CLIP-embedded candidate image feature and viewpoint embedding~\cite{zero123} to the Zero123's diffusion UNet~\cite{zero123}.

\paragraph{Training 3D adapter with proxy samples.}
To train the 3D adapter with coarse proxy as conditions, we preprocess each training object item into several multiview images and uniformly sampled points on the surface.
For each training step, we randomly select $B$ conditioning and target images with the corresponding point samples, and also $B$ timestamp with the Gaussian noise $\epsilon^{(1:B)} \sim \mathcal{N}\in(0, 1)$.
We enforce the network to predict the added noises following~\cite{ddim, stable_diffusion}, which is defined as:
\begin{equation}
    \min_{\theta}\mathbb{E}_{t,\textbf{x}_{(1:N_v)},\epsilon_{(1:N_v)}} \|\epsilon_i-\epsilon_{\theta}(\textbf{x}^t_i,t,c(I,F_C^t,\textbf{c}_i))\|,
\end{equation}
where $\epsilon_{\theta}$ is the model predicted noise, $c(I,F_C^t,\textbf{c}_i)$ is the conditioned embedding of candidate image $I$, 3D control volume $F_C^t$ and camera view $\textbf{c}_i$.
During the training procedure, we use zero convolution~\cite{controlnet} for the proxy feature volume convolution UNet $f_{\text{VP}}$ while freezing other layers, which enables manipulating control strength during the generation.

\subsection{Interactive Generation Workflow}
\label{ssec:method_interact}

In 3D modeling tasks, artists are likely to adjust the target object back and forth, and progressively edit the local part for satisfactory results.
However, the interactive generation and previewing for 3D objects remains an open problem due to the lack of fine-grained controlling ability and slow reconstruction speed~\cite{li2023focaldreamer,cheng2023progressive3d}.
Hence, we develop a novel interactive and responsive generation workflow upon the Coin3D framework, which fully leverages the piecewise proxies of the condition for easy and precise part editing, and reuses 3D control volume for interactive previewing.

\begin{figure}[t!]
    \centering
    \includegraphics[width=1.0\linewidth, trim={0 0 0 0}, clip]{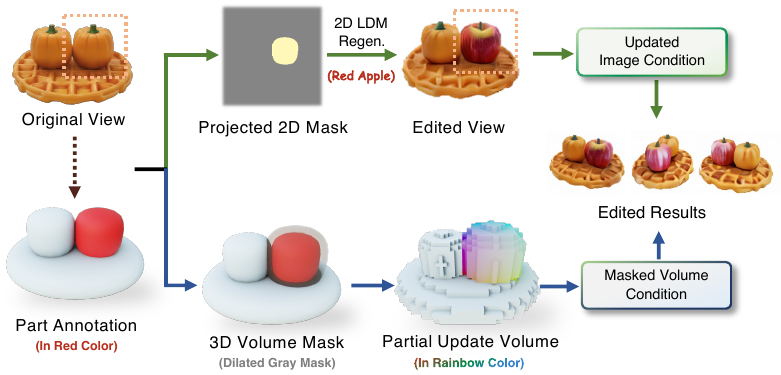}
    \caption{
    \textbf{Proxy-bounded part editing.}
    We update the 2D image condition and 3D control volume with masks from users' part annotation of the proxy.
    }
    \label{fig:local_edit}
\end{figure}

\paragraph{Proxy-bounded part editing.}
As the coarse proxies are mainly constructed with basic shape elements, we design an interactive local part editing workflow based on the elements in the proxy.
Specifically, users can specify a certain piece from the basic shapes, and regenerate the piece content.
For example, we can regenerate one of the pumpkins into a red apple by designating the sphere on the plate, as shown in Fig.~\ref{fig:local_edit}.
However, because the multiview diffusion model is both conditioned on 3D volume and 2D images, it is not trivial to realize the editing regardless of the complete conditions.
Therefore, we propose a two-pathway condition editing scheme that considers both 2D and 3D conditions, as illustrated in Fig.~\ref{fig:local_edit}.
For 2D conditions, we construct a 2D mask by projecting masked proxies at the desired editing view and perform diffusion-based 2D regenerating (a.k.a. masked image-to-image inpainting)~\cite{zhou2023repaint,meng2021sdedit} with the mask.
We then use the edited image as the image condition for the denoising steps.
For 3D conditions, we first construct a 3D feature mask by slightly dilating the masked proxy, which ensures seamless fusion of the newly generated content.
Then, during each denoising step, we reuse the cached original 3D control volume and only partially update the unmasked volume according to the feature mask $M$, as:
\begin{equation}
    \hat{F}_C^t=(1-M)F_C^t + M \tilde{F}_C^t,
\end{equation}
where $\hat{F}_C^t$ is the updated volume by fusing cached volume $F_C^t$ and predicted volume $\tilde{F}_C^t$ at $t$.
By enabling proxy-bounded masks on both 2D and 3D conditions, we can precisely edit the local part at the original object while preserving other parts unchanged.

\paragraph{Interactive preview with progressive volume caching.}
To ensure a smooth experience for interactive generation, we want to preview the editing results in a few seconds and inspect the edited effect from arbitrary viewpoints.
Hence, we design a progressive volume caching mechanism, which memorizes the latest 3D control volume for each timestamp $t$.
Then, during the preview stage, we transfer the user's viewpoint spanning poses $\textbf{c}'$ inside the modeling software to the viewpoint condition and volume projection in multiview diffusion.
To make the preview responsive, we use the cached 3D control volume without re-running the 3D adapter, and instantly decode~\cite{taesd} the preview image for each step.

\subsection{Volume-Conditioned Reconstruction}
\label{ssec:method_recon}

The outcome of the diffusion model is a set of multiview images of the object, so we need to reconstruct it to 3D representation (\eg, using NeuS~\cite{neus}) for CG applications.
However, na\"ively reconstructing with multiview images is sub-optimal and might result in unexpected geometry due to limited viewpoints (see Sec.~\ref{ssec:exp_ablation}).
Therefore, we integrate 3D-aware context from the 3D control volume $F^t_C$ to the reconstruction stage, which improves the reconstruction quality.
Specifically, we propose a volumetric-based score distillation sampling, called volume-SDS, 
which integrates the 3D control prior from the voxelized feature $F^t_C$ to the field's back-propagation as the following:
\begin{equation}
    \Delta_x L_{V-SDS}=w(t)(\epsilon_{\theta}(\textbf{x}_t,t,c(I,F_C^t,\textbf{c}))-\epsilon),
\end{equation}
where $w(t)$ is the weighting function~\cite{dreamfusion}.
In this way, the reconstruction can be decently supervised by 3D control signals to achieve better mesh quality (see Sec.~\ref{ssec:exp_ablation}).
Please refer to the the supplementary material for more details of the reconstruction.

\section{Experiments}

We first compare our method with image-based 3D generation in Sec.~\ref{ssec:compare_mv}, and compare with controllable 3D object generation methods in Sec.~\ref{ssec:compare_3d_control}.
Then, we show the interactive generation applicability with designated part editing in Sec.~\ref{ssec:app_part_edit}.
Finally, we perform ablation studies to analyze the design of
our framework in Sec.~\ref{ssec:exp_ablation}.

\begin{figure*}[t!]
    \centering
    \includegraphics[width=1.0\linewidth, trim={0 0 0 0}, clip]{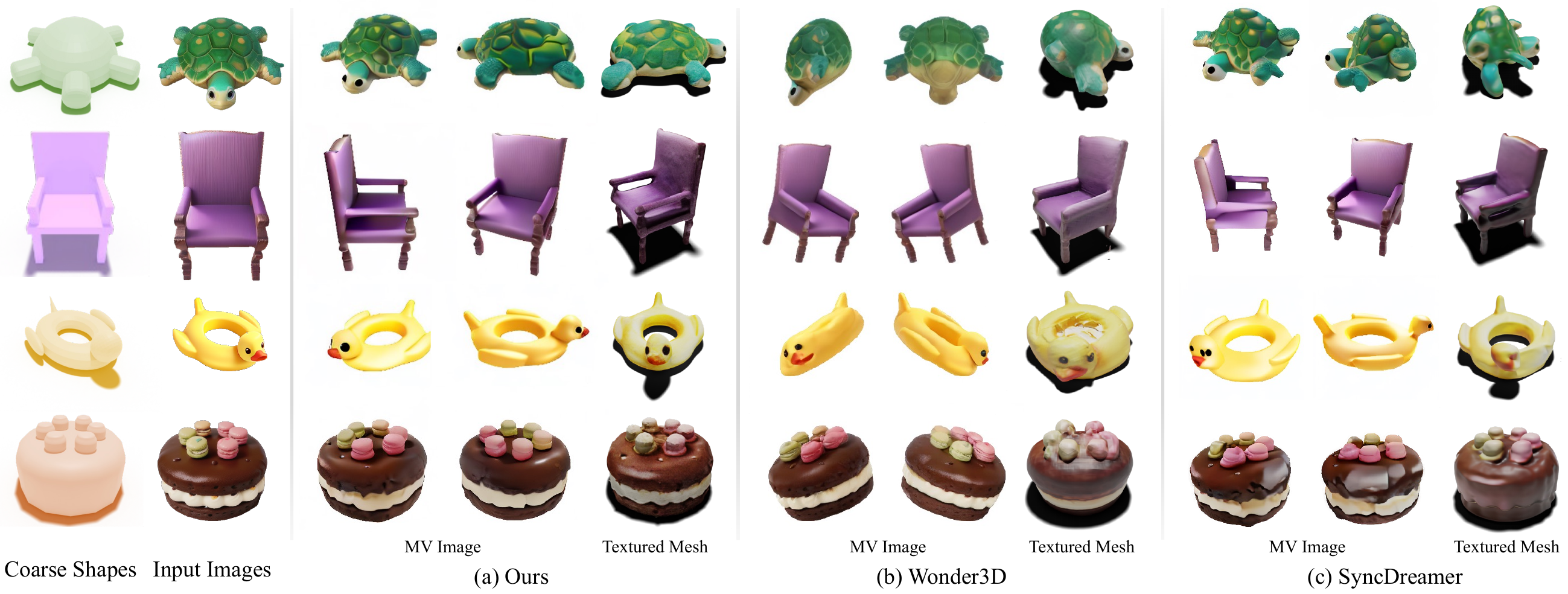}
    \caption{
    We compare our proxy-based generation method with image-based methods (\ie, Wonder3D~\cite{wonder3d} and SyncDreamer~\cite{syncdreamer}) on the generated multiview images and reconstructed textured mesh.
    }
    \label{fig:exp_multiview_mesh}
\end{figure*}

\begin{table}[t!]
\centering
\resizebox{1.0\linewidth}{!}{
\tabcolsep 2pt
\begin{tabular}{lcccccc}
\toprule
\multicolumn{1}{c}{\multirow{2}{*}{Methods}} & \multicolumn{3}{c}{Quantitative Metrics} & \multicolumn{2}{c}{User Study} \\ \cmidrule(lr){2-4} \cmidrule(lr){5-6}  
\multicolumn{1}{c}{} & \multicolumn{1}{l}{CLIP Score $\uparrow$} & \multicolumn{1}{l}{ImageReward $\uparrow$} & \multicolumn{1}{l}{GPTEvals3D $\uparrow$} & \multicolumn{1}{l}{Matching Degree $\uparrow$} & \multicolumn{1}{l}{Recon. Quality $\uparrow$} \\ 
\hline
\multicolumn{6}{c}{Proxy-Based vs. Image-Based 3D Generation} \\ 
% \hline
\hdashline
% \midrule
Wonder3D~\cite{wonder3d} & 0.251 & -0.557 & 
 980 & 1.613 & 1.770 \\
SyncDreamer~\cite{syncdreamer} & 0.260 & -0.152 & 962 & 1.654 & 1.594 \\
Ours & \textbf{0.266} & \textbf{0.026} & 
\textbf{1035} & \textbf{2.733} & \textbf{2.634} \\
\hline
\multicolumn{6}{c}{Controllable 3D Object Generation} \\ 
% \hline
\hdashline
Fantasia3D~\cite{fantasia3d} & 0.212  & -1.597 & 810 & 1.267 & 1.273 \\
Latent-NeRF~\cite{latentnerf} & 0.246 & -1.188 & 1146 & 1.930 & 1.918 \\
Ours & \textbf{0.249} & \textbf{-0.749} & 
\textbf{1204} & \textbf{2.801} & \textbf{2.809} \\
\bottomrule
\end{tabular}
}
\newline
\newline
\caption{
We perform quantitative evaluation and user studies on the 3D generation task.
}

\label{tab:compare}
\end{table}

\subsection{Comparison on Proxy-based and Image-based 3D Generation}
\label{ssec:compare_mv}

So far, the most stable 3D object generation pipelines are mainly image-based, \ie, giving a single image as a conditioning input, and then generating multiview images for reconstruction~\cite{zero123++,wonder3d,syncdreamer} or direct 3D representations~\cite{lrm}.
Unlike these methods, we use a coarse shape proxy as a guidance through the entire interactive generation pipeline.
Since all these methods use image conditions to bootstrap the diffusion model, we first compare our method with SOTA image-based generation methods (\ie, Wonder3D~\cite{wonder3d} and SyncDreamer~\cite{syncdreamer}) using the same image candidates, where our method also add extra coarse shapes as conditioning.

\paragraph{Qualitative comparison.}
We show the multiview images and the reconstructed textured meshes in Fig.~\ref{fig:exp_multiview_mesh}.
As shown in Fig.~\ref{fig:exp_multiview_mesh}, the predicted views and the textured meshes from Wonder3D and SyncDreamer both have some artifacts (\eg, distorted green turtle and yellow swimming ring at the first and third row in Fig.~\ref{fig:exp_multiview_mesh} (b) (c), missing hollowed handrail and short legs at the second row in Fig.~\ref{fig:exp_multiview_mesh} (b), missing white creamy middle layer at the fourth row in Fig.~\ref{fig:exp_multiview_mesh} (c)).
Thanks to the proxy-guided conditioning and volume-conditioned reconstruction, our method can synthesize multiview images free of single view ambiguity by complementing 3D context from the proxy (\eg, complete chairs with correct hollowed handrail in Fig.~\ref{fig:exp_multiview_mesh} (a)), and also consistently reconstruct 3D objects with intact shape and vivid appearance.

\paragraph{Quantitative comparison.}
We use CLIP score~\cite{radford2021learning} to evaluate the text-object matching degree, and ImageReward~\cite{xu2023imagereward,he2023t} and GPTEvals3D~\cite{wu2024gpt} to evaluate the perceptual quality of the predicted multiview images.
As presented in Table~\ref{tab:compare},
our method achieves the overall best metrics, demonstrating that adding proxy-based conditioning can improve the quality of 3D generation tasks.
Note that Wonder3D's ImageReward score is lower than SyncDreamer's due to the evaluator's bias of orthogonal image views, while their Elo scores~\cite{elo1967proposed} evaluated by GPTEvals3D are comparable.

\paragraph{User study.}
We also conduct a user study to compare our method with others.
Following TEXTure~\cite{texture_paper}, we ask 30 users to sort 35 testing examples in random order based on the perceptual quality and content matching degree (w.r.t the given image or text prompts), and assign the scores by their ranking (\ie, with a score of 3 for the ordered best one and a score of 1 for the last one).
As reported in Table~\ref{tab:compare}, our method achieves the best score among all the methods.
More details can be found in the supplementary material.

\begin{figure*}[t!]
    \centering
    \includegraphics[width=1.0\linewidth, trim={0 0 0 0}, clip]{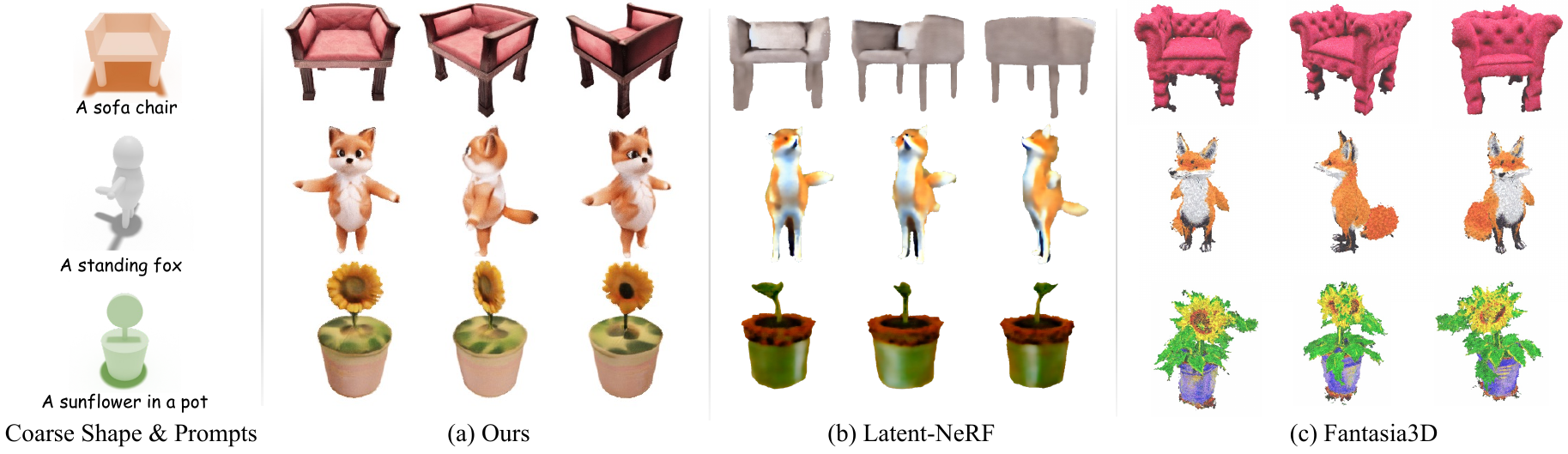}
    \caption{
    We compare the Controllable 3D  generation with Latent-NeRF~\cite{latentnerf} and Fantasia3D~\cite{fantasia3d}.
    }
    \label{fig:exp_3d_control}
\end{figure*}

\begin{figure*}[t!]
    \centering
    \includegraphics[width=1.0\linewidth, trim={0 0 0 0}, clip]{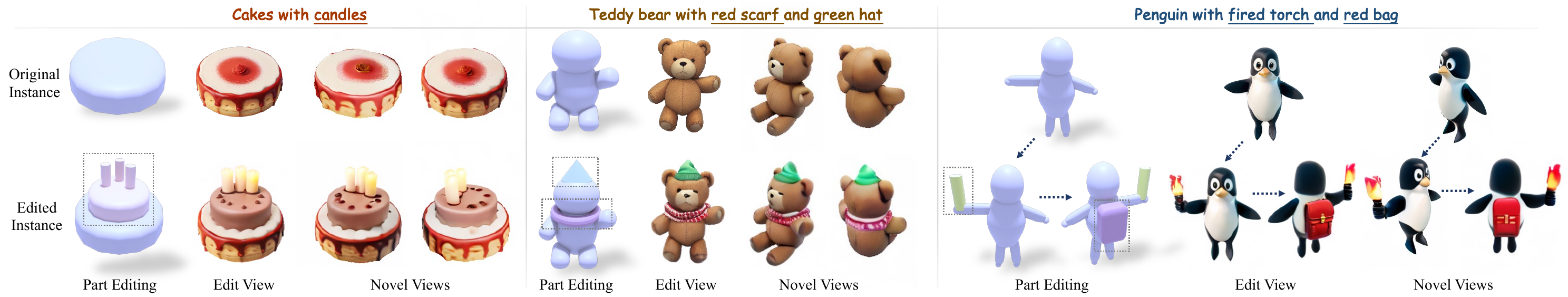}
    \caption{
    We conduct interactive generation with part editing on several basic shape proxies.
    }
    \label{fig:exp_interactive}
\end{figure*}

\subsection{Comparison on Controllable 3D Object Generation}
\label{ssec:compare_3d_control}

We compare our method with Controllable 3D object generation methods, including Latent-NeRF~\cite{latentnerf} and Fantasia3D~\cite{fantasia3d}.
Latent-NeRF introduces a sketch-shape guided loss, which constrains the density field close to the surface of the shape proxy, while Fantasia3D uses coarse shapes as the geometry initialization of DMTet~\cite{dmtet}.
In the experiment, we give all the methods the same coarse shape and the text prompts as a guidance and output the neural reconstructions' rendered views for comparison (since Latent-NeRF does not officially provide mesh extraction).
As shown in Fig.~\ref{fig:exp_multiview_mesh}, Latent-NeRF generally obtains plausible results but fails to produce tiny shape and appearance details (\eg, blurry textured sofa, fox missing clear eyes and right arms and missing sunflower in Fig.~\ref{fig:exp_3d_control} (b)), which indicates that directly applying 3D control to the 3D representation is sub-optimal since it might not work smoothly with the generic SDS loss.
For Fantasia3D, since it only uses the 3D shape for initialization rather than supervision, it often generates overgrowth results that do not follow the given shape (\eg, inflate sofa, the fox with incomplete downward arms, pot with many leaves and a broken sunflower in Fig.~\ref{fig:exp_3d_control} (c)) and also slightly suffers from ``multi-face Janus problem'' (\eg, multi-face fox in Fig.~\ref{fig:exp_3d_control} (c)).
Since both Latent-NeRF and Fantasia3D use vanilla 2D diffusion model as prior while being agnostic to the multi-view correlations, their results are sensitive to the initialization and hyperparameter settings.
In contrast, our method directly adds 3D-aware control to the diffusion process, which essentially controls the supervisory of reconstruction's 2D diffusion prior and consistently achieves high fidelity generation following users' shape guidance.
It is also noteworthy that, both Latent-NeRF and Fantasia3D require a long period of reconstruction (\eg, dozens of minutes) to give an impression of what the object might look like, making it unusable for interactive modeling, while our framework bypasses the reconstruction stage and allows to preview the 3D object in only a few seconds.

\subsection{Interactive Generation with Part Editing}
\label{ssec:app_part_edit}

We now present examples of interactive generation with progressive part editing.
As shown in Fig.~\ref{fig:exp_interactive}, users can first generate a basic instance (\eg, a base of cake, a teddy bear, or a penguin) with shape proxy, and then progressively add new shape blocks with changed text prompts (\eg, adding a small cake with candles, a green hat and red scarf, or even progressively add a torch and a red backpack from the back view), which seamlessly enrich the content of the instance while maintaining other parts unchanged.
Notably, all these editing operations can be finished in roughly 5$\sim$10 seconds, which then allows interactive previewing of the edited 3D results.
Please refer to the supplementary video for the demonstration.

\begin{figure}[t!]
    \centering
    \includegraphics[width=1.0\linewidth, trim={0 0 0 0}, clip]{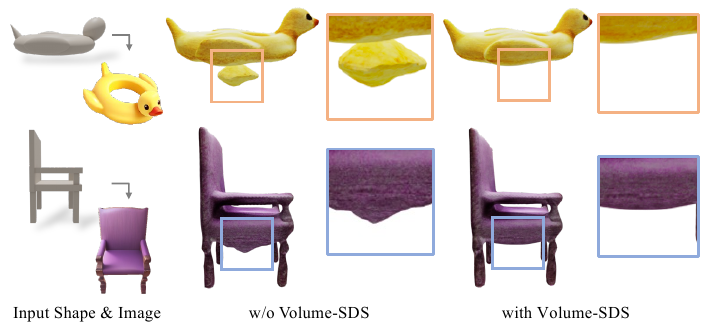}
    \caption{
    We inspect the efficacy of volume-conditioned reconstruction.
    }
    \label{fig:ablation_volume_sds}

\end{figure}

\begin{figure}[t!]
    \centering

    \includegraphics[width=1.0\linewidth, trim={0 0 0 0}, clip]{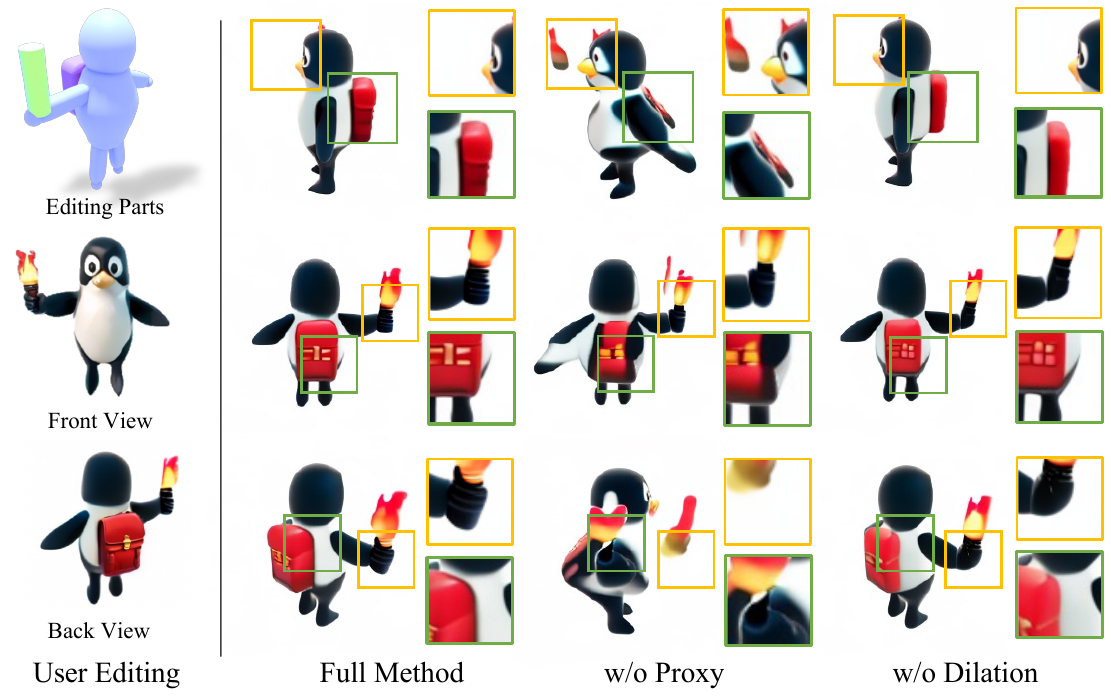}
    \caption{
    We analyze the importance of proxy guidance and 3D mask dilation in proxy-bounded part editing.
    }
    \label{fig:ablation_seamless}

\end{figure}

\subsection{Ablation Studies}
\label{ssec:exp_ablation}

\paragraph{Volume-conditioned reconstruction.}
We then inspect the efficacy of the volume-SDS loss by ablating during the shape reconstruction.
As shown in Fig.~\ref{fig:ablation_volume_sds}, by adding volume-SDS loss, we achieve better geometry reconstruction (\eg, less floater and more reasonable chair bottom) than na\"ive training on fixed multiview images~\cite{wonder3d,syncdreamer}.

\paragraph{Proxy-bounded part editing.}
We finally analyze the proxy-bounded part editing by ablating proxy conditioning and mask dilation strategy in Fig.~\ref{fig:ablation_seamless}.
Specifically, we choose a multi-step editing example, where the backpack should be edited from the back view.
We merge the front and edited back image conditions using mixed denoising~\cite{bar2023multidiffusion}.
As shown in Fig.~\ref{fig:ablation_seamless}, the editing without the proxy condition would result in broken shapes (\eg, dangling flames and distorted body), while disabling mask dilation would also make the editing less natural (\eg, tightly fused red bag and broken hands).
By equipping with the full strategies, we achieve a seamless editing effect while preserving other content unchanged.

More ablation studies can be found in the supplementary material.
\section{Conclusions}

We have proposed a novel 3D assets generation pipeline, named Coin3D.
Our method successfully integrates 3D-aware control from coarse shape proxies to the 3D object generation task and enables an interactive generation workflow, where users can freely alter prompts/shapes or regenerate designated local parts, and inspect the changes with interactive 3D preview in a few seconds.

\paragraph{Limitations and future works.}
First, our workflow starts from synthesizing 2D image candidates, which provides users with quick preview and selection but requires prompt engineering to obtain a clean and satisfactory result without complex background textures.
In the future, we can finetune a 2D diffusion model with object-centric data~\cite{objaverse} and introduce LLM-based prompt enhancement~\cite{magicprompt} to handle this issue.
Second, due to the limited resolution of the base diffusion model~\cite{zero123}, our method cannot produce fine-level details (\eg, complex fur textures or wrinkled surface), which can be further improved by adopting better backbones~\cite{zero123++} or taking the refinement stage with high-resolution optimization~\cite{magic3d,tang2023dreamgaussian}.
Third, our reconstructed texture meshes already baked lighting effects while lacking PBR materials for modern rendering pipelines. 
In future work, we can train a material-disentangled diffusion model to enable generating objects with PBR materials.

\begin{acks}
We would like to acknowledge the support from the NSFC (No.~62102356), Information Technology Center and State Key Lab of CAD\&CG, Zhejiang University. 
We also express our gratitude to all the anonymous reviewers for their professional and constructive comments. 
The authors from Zhejiang University are also affiliated with the State Key Laboratory of CAD\&CG.
\end{acks}
% \clearpage

% Bibliography
\bibliographystyle{ACM-Reference-Format}
\bibliography{main}

\clearpage

\appendix

\renewcommand\thesection{\Alph{section}}
\renewcommand\thetable{\Alph{table}}
\renewcommand\thefigure{\Alph{figure}}

{\noindent \Large \bf Supplementary Material\par}

\vspace{1em}

In this supplementary material, we describe more details of our method in Sec.~\ref{sec:impl}.
Besides, we also conduct more experiments in Sec.~\ref{sec:more_expr}.
More qualitative results can be found in our supplementary video, and the source code will be released upon the acceptance of this paper.

\section{Implementation Details}
\label{sec:impl}

\subsection{Dataset Preparation}

In our experiment, we use the LVIS subset of Objaverse~\cite{objaverse} to train the model, which contains 28,000+ objects after a heuristic cleanup process following Long \etal~\cite{wonder3d}.
For training view rendering, we set up 16 image views with -30° pitch and evenly facing towards the object from 360°.

\subsection{Evaluation Data Preparation and User Study}
For quantitative comparison, we produced 30 testing examples (coarse shapes and users' prompts) for each experiment (Sec. 4.1 and Sec. 4.2). Then, for each example, we generate four images at four poses $\{\textbf{c}_i|i= 1, 5, 9, 13\}$ from 16 evenly distributed viewpoints, and calculates CLIP score~\cite{radford2021learning}, ImageReward~\cite{xu2023imagereward,he2023t} and GPTEvals3D~\cite{wu2024gpt} average Elo scores for each standalone view.
For user studies, we prepare 35 examples and merge the output images of each method into one image. Then we ask 30 participants to sort the merged images.
In the comparison on proxy-based and image-based 3D generation, we merged four multiview images and four textured mesh rendering images into one. 
In the comparison on controllable 3D object generation, we merge four rendering images into one, since LatentNeRF is difficult to extract the textured mesh.

\subsection{Training and Network Details}

During the initialization of the training, we follow Zhang\etal~\cite{controlnet} to keep the multiview convolution network weights $f^{\text{VM}}$ fixed, and use a dual 3D UNet structure to implement the 3D feature Adapter with trainable copy initialization strategy.
Specifically, our proxy is first voxelized at a resolution of $ 32 \times 32 \times 32$, where the value of each voxel would be assigned to 1 if there is any occupied 3D point inside the voxel.
After that, the features will be up-convolution to 64 channels through two layers of 3D convolution.
For the training of the 3D adapter, we sample 256 points on each object surface as a coarse proxy, and train the model at $256 \times 256$ resolution. The learning rate is 0.00005 and the batch size is set to 8, with 100K training iterations.
The total training process of the 3D Adapter takes about two days on an Nvidia A100-80G graphic card.

During the textured mesh reconstruction stage, we use NeuS~\cite{neus} as the neural representation. The total loss function is defined as the following:
\begin{equation}
    L=L_{rgb}+L_{V-SDS}+L_{mask}+R_{eik}+R_{sparse}+R_{smooth}
\end{equation}
where $L_{rgb}$ is the $L_2$ loss between the generated multiview images $\textbf{x}_{(i:N_v)}$ and images rendered under the camera poses $\textbf{c}_{(i:N_v)}$.
$L_{V-SDS}$ is the volume-SDS loss proposed in the main paper.
$L_{mask}$ is the BCE mask loss of the rendered opacity, where the mask $M$ is obtained with existing methods~\cite{kirillov2023segment}.
$R_{eik}$ is the Eikonal loss that regularize the magnitude of the SDF gradients of each sample point to be unit length.
$R_{sparse}$ and $R_{smooth}$ are respectively the sparse term used to reduce the floater of SDF and the smooth term used to smooth the 3D surface. 
For the 3D reconstruction of each instance, we takes about 5 minutes on a single NVidia A100-80G graphic card.

\begin{figure}[t!]
    \centering
    \includegraphics[width=1.0\linewidth, trim={0 0 0 0}, clip]{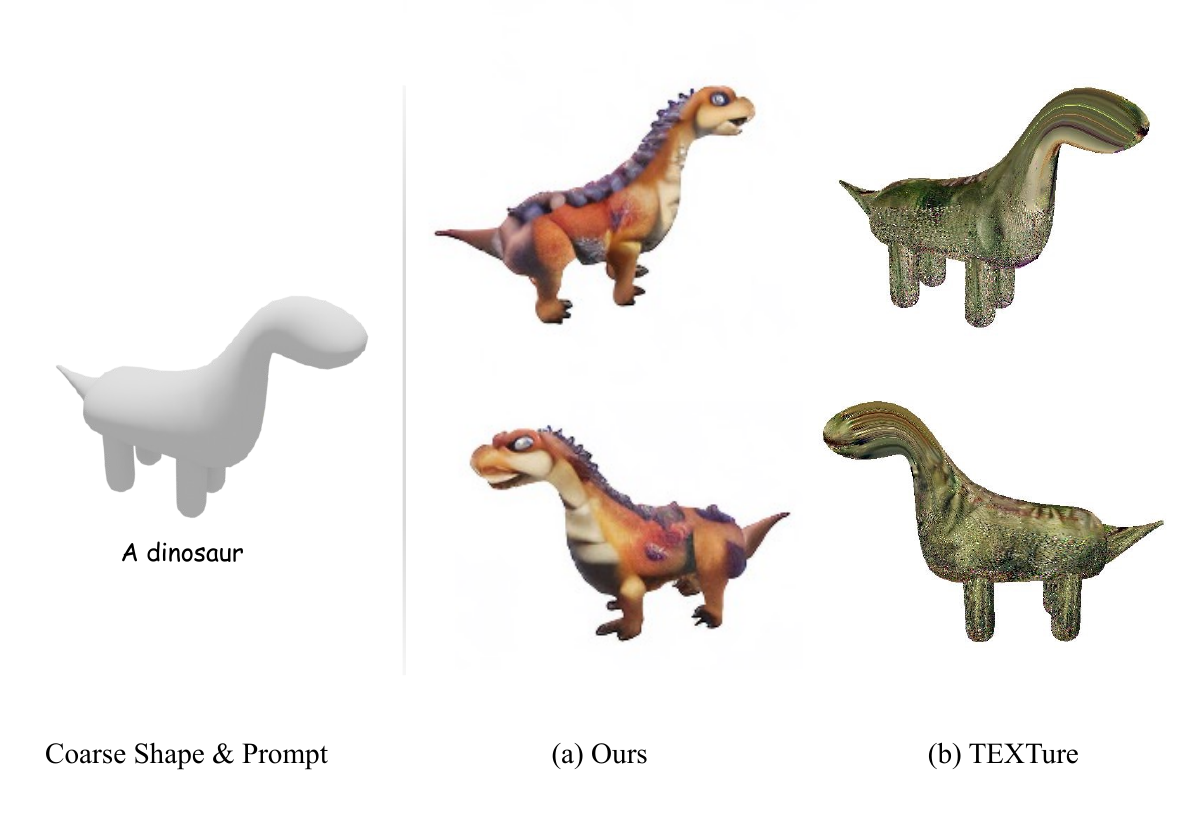}
    \caption{
    We compare our method with the texture synthesis method TEXTure~\cite{texture_paper}.
    }
    \label{fig:supp_texture}
\end{figure}

\begin{figure}[t!]
    \centering
    \includegraphics[width=1.0\linewidth, trim={0 0 0 0}, clip]{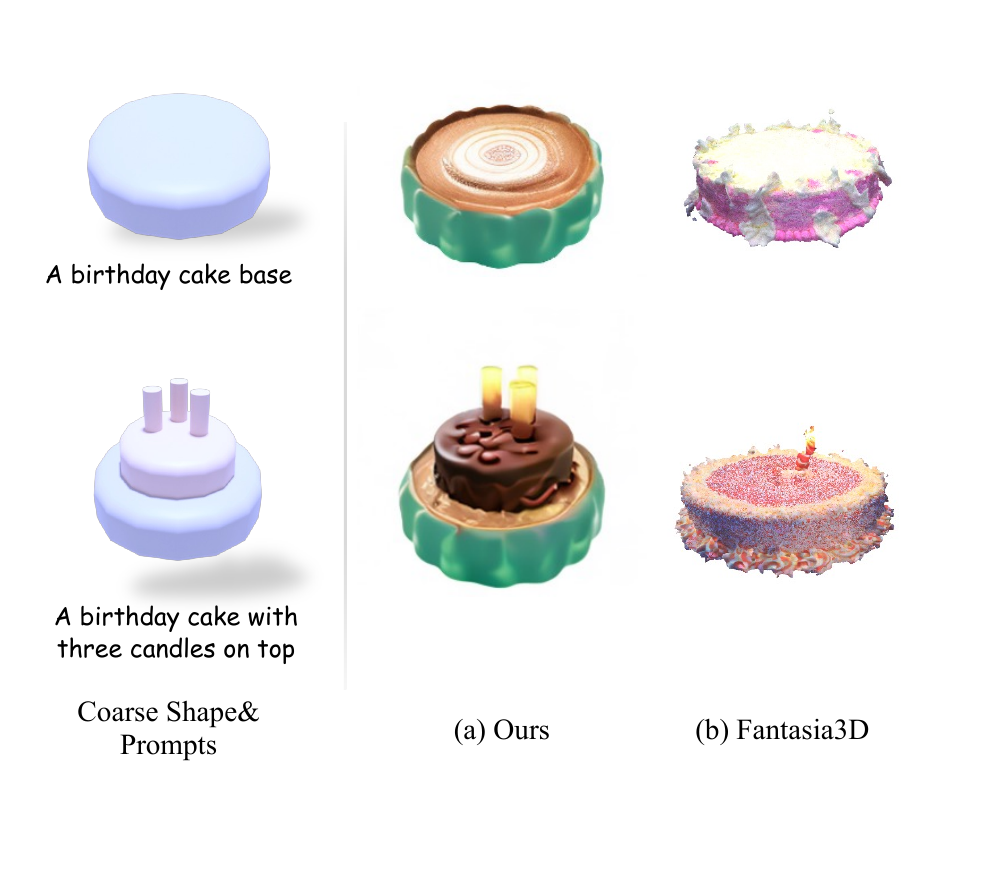}
    \caption{
    We compare our proxy-bounded part editing with Fantasia3D~\cite{fantasia3d} fine-tuning.
    }
    \label{fig:supp_fantasia3d}
\end{figure}

\begin{figure}[t!]
    \centering
    \includegraphics[width=1.0\linewidth, trim={0 0 0 0}, clip]{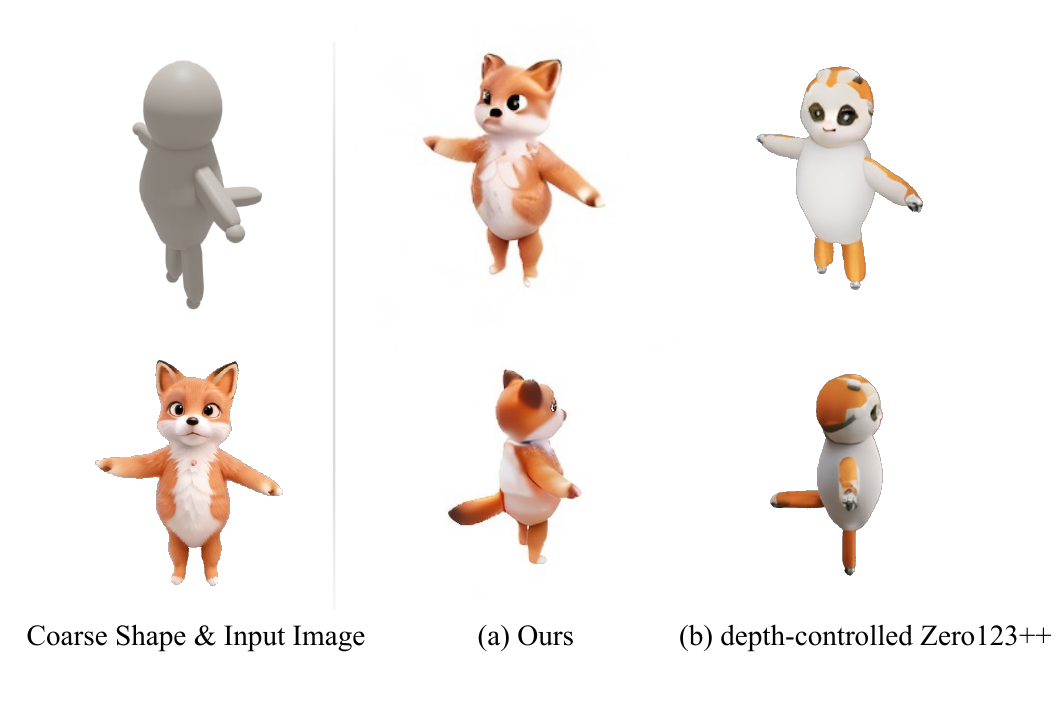}
    \caption{
    We compare our method with the depth-controlled 3D generation pipeline (Zero123++~\cite{zero123++}).
    }
    \label{fig:supp_zero123pp}
\end{figure}

\section{More Discussions}
\subsection{Necessity of Proxy-guided 3D Generation}
For personalized generation demands, we think only using text / images is insufficient and also unintuitive for expressing 3D structures of objects and their spatial relationships. Hence, granting system 3D-aware controllability with 3D proxy is necessary for 3D generation. As for the acquisition of 3D proxies, we believe this is not an obstacle for target users, as it can be assembled easily using kids' software like Tinkercad, taken from 3D modeling games from SteamVR, or using LLM+procedural modeling instructions. Similarly, ControlNet uses control images from raw sketches to delicate line art, which also requires basic painting skills. 

\subsection{More Limitations}
First, the resolution of 3D-aware control is bounded by the size of the proxy feature volume, which cannot fully leverage control from complex high-poly models. For example, we cannot generate a large-scale urban scene with satisfactory building details. Second, our method requires manual tuning control strength to balance between over-constrained and under-constrained, which is also similar to ControlNet~\cite{controlnet} as the control strength mainly depends on the creators' aesthetic choices.

\section{More Experiments}
\label{sec:more_expr}

\begin{figure}[t!]
    \centering
    \includegraphics[width=1.0 \linewidth, trim={0 0 0 0}, clip]{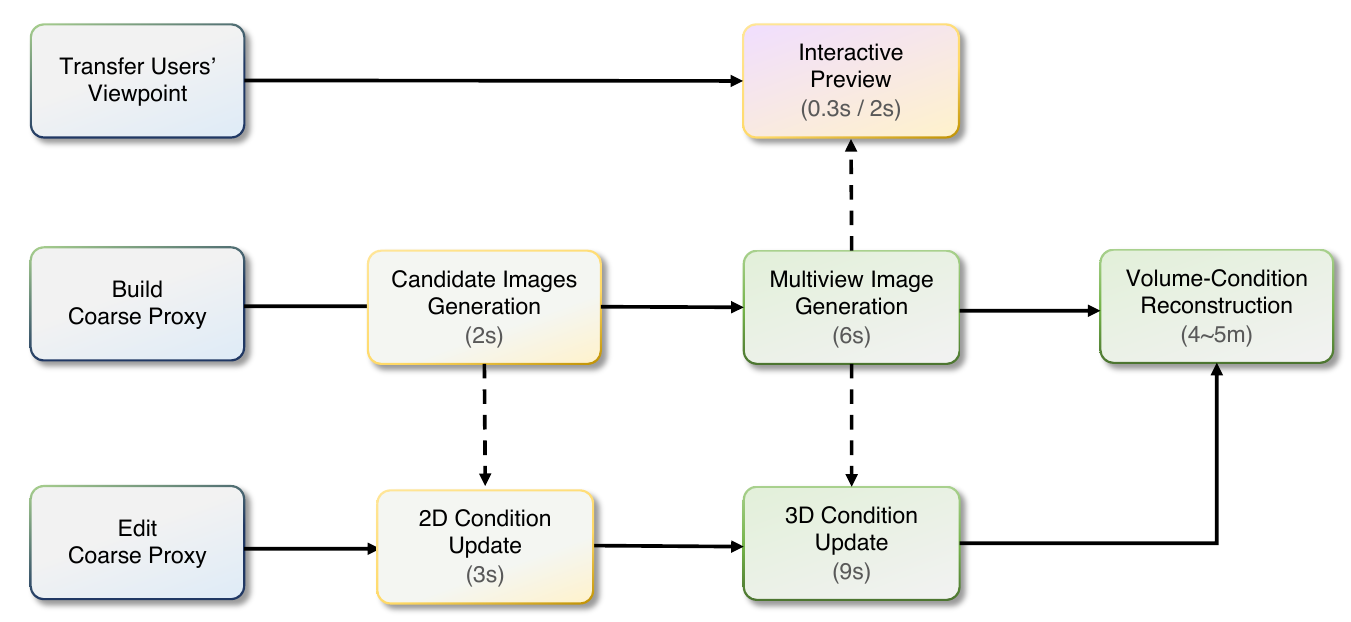}
    \caption{
    Runtime overview of interactive generation.
    }
    \label{fig:supp_more_exp_runtime}
\end{figure}

\subsection{Runtime Evaluation}
As shown in Fig.~\ref{fig:supp_more_exp_runtime}, given a 3D proxy and text prompt, Coin3D first takes 2s to generate candidate images and 6s to generate 3D-aware conditioned multiview images. During previewing, Coin3D responds to the camera rotating instantly (<0.3s) and takes 2s for convergence (baseline without volume-cache requires 6s). For interactive editing, Coin3D takes 3s to update the 2D condition and 9s to update the feature volume, which is then ready for previewing. As a comparison, existing editable 3D generation methods take much longer for feedback, e.g., \textasciitilde 1h for Progressive3D and FocalDreamer, \textasciitilde 0.5h for Fantasia3D. Finally, Coin3D takes 4\textasciitilde5m (600 iterations) to reconstruct and export the textured mesh (see Fig. 1 and Fig. 4 from the main paper for "proxies vs. textured meshes").

\begin{figure}[t!]
    \centering
    \includegraphics[width=1.0\linewidth, trim={0 0 0 0}, clip]{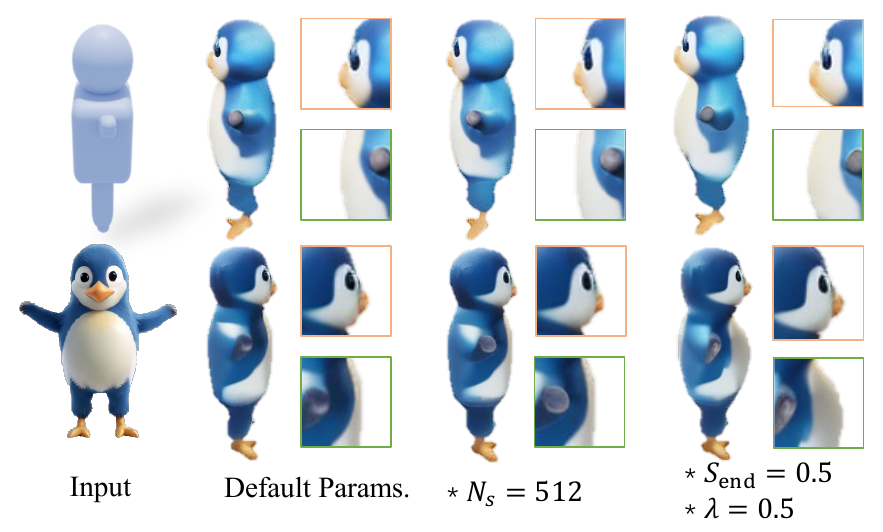}
    \caption{
    We analyze the 3D-aware controlling strength with different control parameters. As shown above, increasing proxy samples would add more shape constraints, while setting lower weights would give more freedom of generation.
    }
    \label{fig:ablation_strength}
\end{figure}

\begin{figure*}[t!]
    \centering
    \ContinuedFloat
    \captionsetup{format=cont}
    \includegraphics[width=1.0 \linewidth, trim={0 0 0 0}, clip]{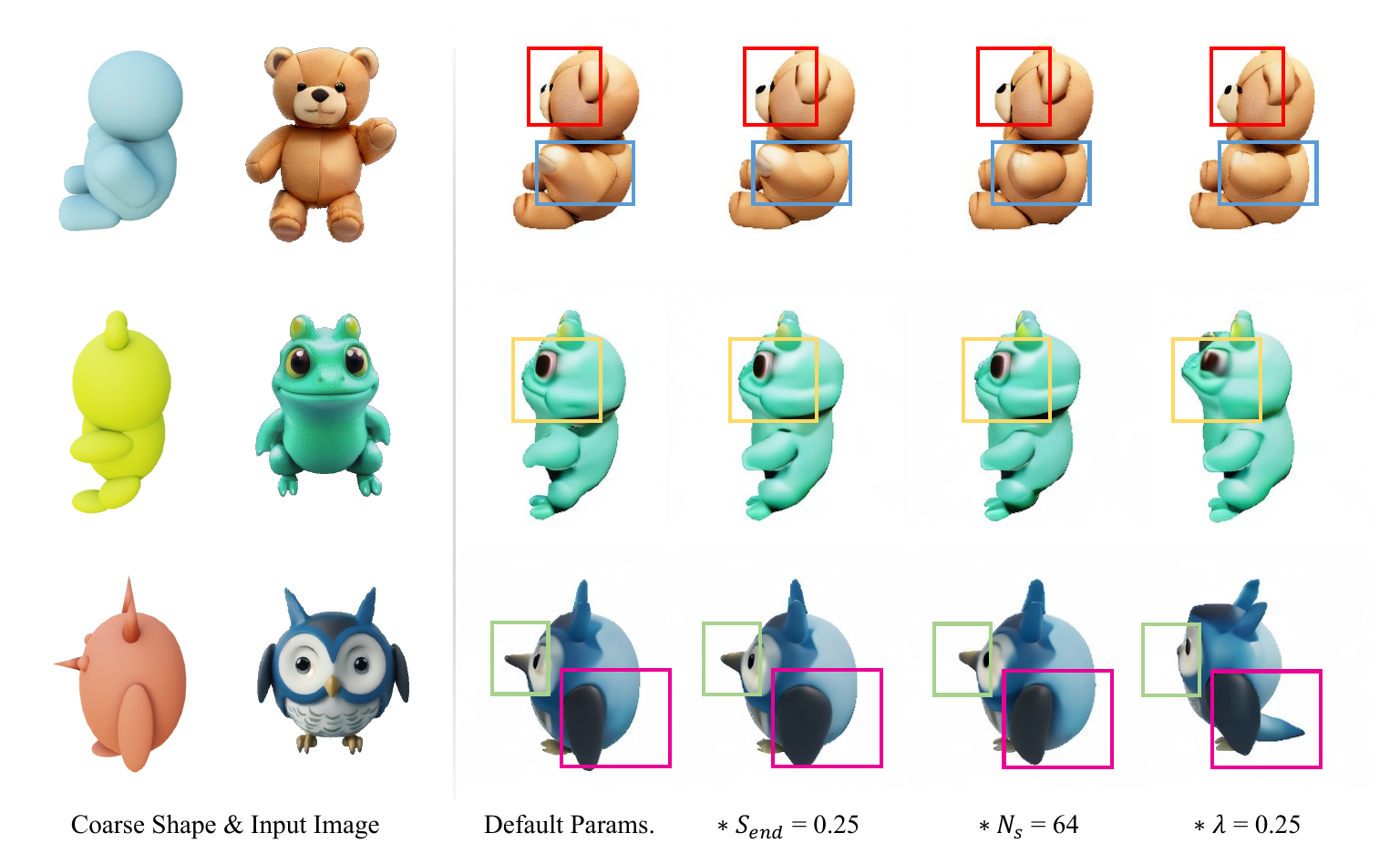}
    \caption{
        We analyze the 3D-aware controlling strength with different control parameters. As shown above, setting lower weights would give more freedom of generation.
    }
    \label{fig:ablation_strength_cont}
\end{figure*}

\subsection{3D-Aware Controlling Strength}

We analyze the adjustable 3D-aware controlling strength with different control parameters under the fixed seed in Fig.~\ref{fig:ablation_strength} and Fig.~\ref{fig:ablation_strength_cont}.
For the default parameters, we set the number of proxy samples $N_s=256$ and fully unlocked weight control ($\lambda=1.0$ and $S_{\text{end}}=1.0$) through the whole diffusion process.
As a comparison, we set the $N_s=512$ or set partial weight control ($\lambda=0.5$ and $S_{\text{end}}=0.5$, \ie, only half weight applied and disable weights for the last half of the denoising steps), where $\lambda$ is the weight when adding intermediate outputs of $f_{VP}$ to $f_{VM}$, and $S_{\text{end}}$ is the ending step of 3D-aware controlling.
We found that increasing the number of proxy samples would add more constraints to make it close to the given shape, while setting lower weights would give more freedom to the network to predict a curved shape.

\subsection{Interactive Editing vs. Fine-Tuning Fantasia3D}

Since Fantasia3D supports fine-tuning on the pre-trained mesh, it can also conduct interactive editing by updating text prompts.
Therefore, we compare our interactive editing method with Fantasia3D fine-tuning.
We use the user-guided generation method mentioned in Fantasia3D. The first stage performs geometry optimization and texture optimization based on the initial shape, and the second stage uses the first stage's optimized object but modifies the text prompt during optimization. 
During the experiment, we first use a flat cylinder as the initial input shape, and use text prompt ``a birthday cake base'' to guide the optimization.
Second, we add candles by modifying the proxy shape and updating the text prompt to ``a birthday cake with three candles on top'' and  
As shown in Fig.~\ref{fig:supp_fantasia3d}, our method successfully generates appealing cakes with candles, while Fantasia3D fails to achieve a reasonable result (\eg, almost no complete candle on the top).

\subsection{Proxy-Based 3D Generation vs. Texture Synthesis}

We also compare our method with texture synthesis work TEXTure~\cite{texture_paper}, which generates UV textures given the corresponding geometry.
As shown in Fig.~\ref{fig:supp_texture}, when given the same coarse proxy for generation, TEXTure tends to generate tightly bounded textures on the given mesh, resulting in blurry appearances and invisible facial features of the dinosaur.
In contrast, our method allows a certain degree of freedom during the generation, which gracefully synthesizes the dinosaur with vivid facial details.
The experiment demonstrates that the proxy-based 3D generation is far beyond the texture synthesis task, as it requires the method to generate more details upon the coarse proxy shape.

\subsection{Proxy-Based 3D Generation vs. Depth-Controlled 3D Generation}

We compare our proxy-based 3D generation with the depth-controlled 3D generation pipeline from Zero123++~\cite{zero123++}, where we feed the Zero123++ with the multiview depth maps rendered from the coarse shape proxy.
As shown in Fig.~\ref{fig:supp_zero123pp}, even only given a coarse shape of the animal with no ears, our method still generates cute animal ears upon the simple shape, while Zero123++ can only synthesize novel views of the object that tightly fit to the coarse shape proxy with poor facial details.
This demonstrates that simply using 2D depth control as a condition in multiview generation cannot achieve the ideal coarse 3D control ability like ours, which further proves the value of adding 3D-aware control in a 3D manner.

\subsection{User Studies of Ablation Studies}
We selected 10 examples of the ablation studies in Sec. 4.4 and asked 24 users to judge whether the proposed strategies improve the results. The statistic shows 78\% of users believe volume-SDS improves the quality, 75\% of users think 3D mask dilation makes editing more natural, 96\% of users find the proxy helpful in maintaining shape integrity.

\subsection{User Study of 3D Interaction of Proxy-based 3D Generation}
We also conducted a user study of proxy-based 3D generation. We show users the process of making coarse shapes in Blender, as well as the generated multi-view images and 3D reconstruction results. We asked each participant to rate three questions: (a) the difficulty of using 3D modeling tools; (b) overall satisfaction with the effectiveness of our approach; (c) willingness to use our methods, on a scale of 1 to 5, where 1 in (a) means easy to use, 5 in (b) means satisfied with the effectiveness, and 5 in (c) means willing to use. The score of (a) is 2.38, (b) is 4.62, and (c) 4.46, which indicates that most of the participants consider the difficulty of coarse shape modeling is acceptable and are willing to use our method.

\begin{figure*}[t!]
    \centering
    \includegraphics[width=1.0 \linewidth, trim={0 0 0 0}, clip]{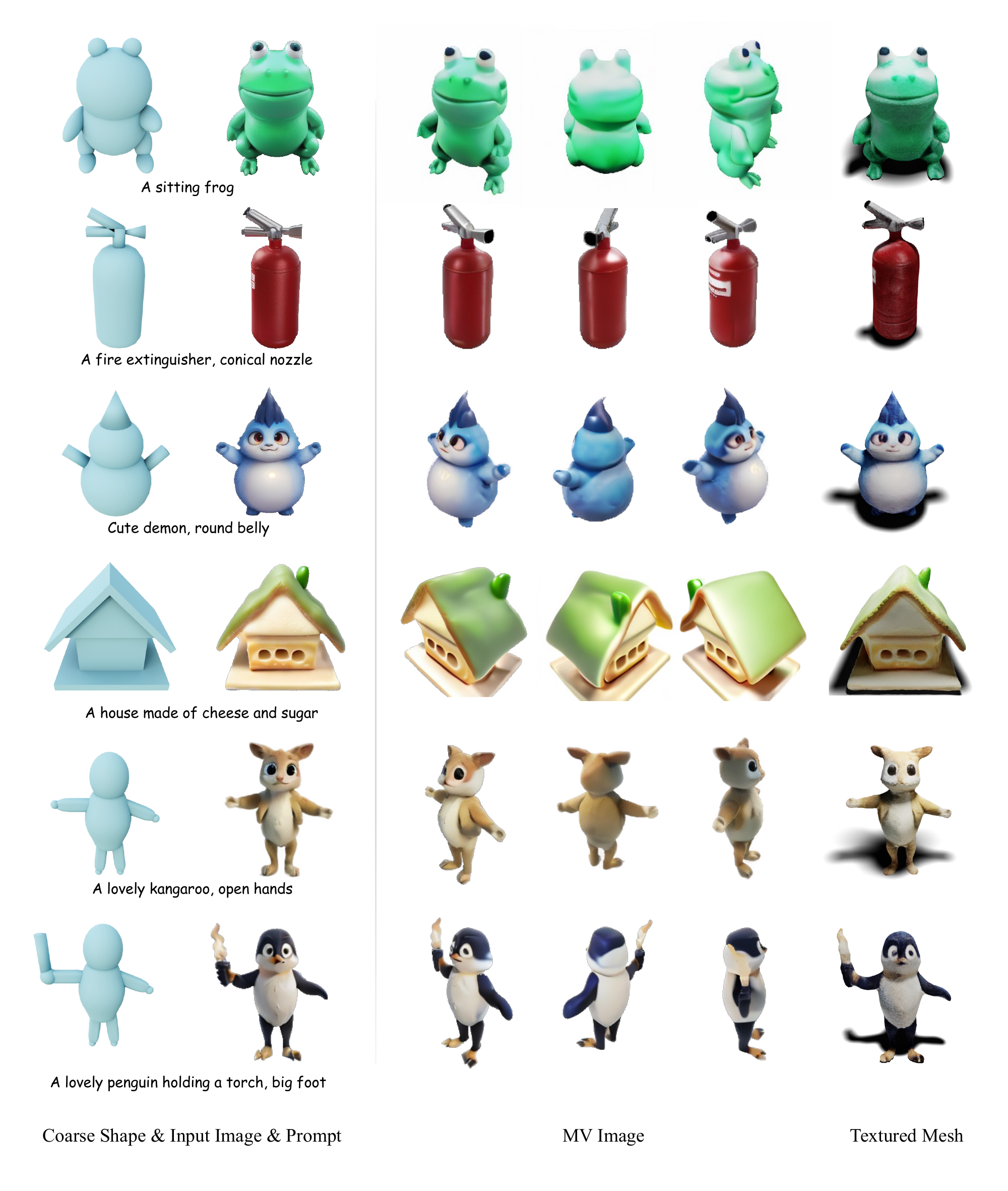}
    \caption{
    More examples of controllable 3D object generation.
    }
    \label{fig:supp_more_exp1}
\end{figure*}

\begin{figure*}[htbp]
    \centering
    \ContinuedFloat
    \captionsetup{format=cont}
    \includegraphics[width=1\linewidth, trim={0 0 0 0}, clip]{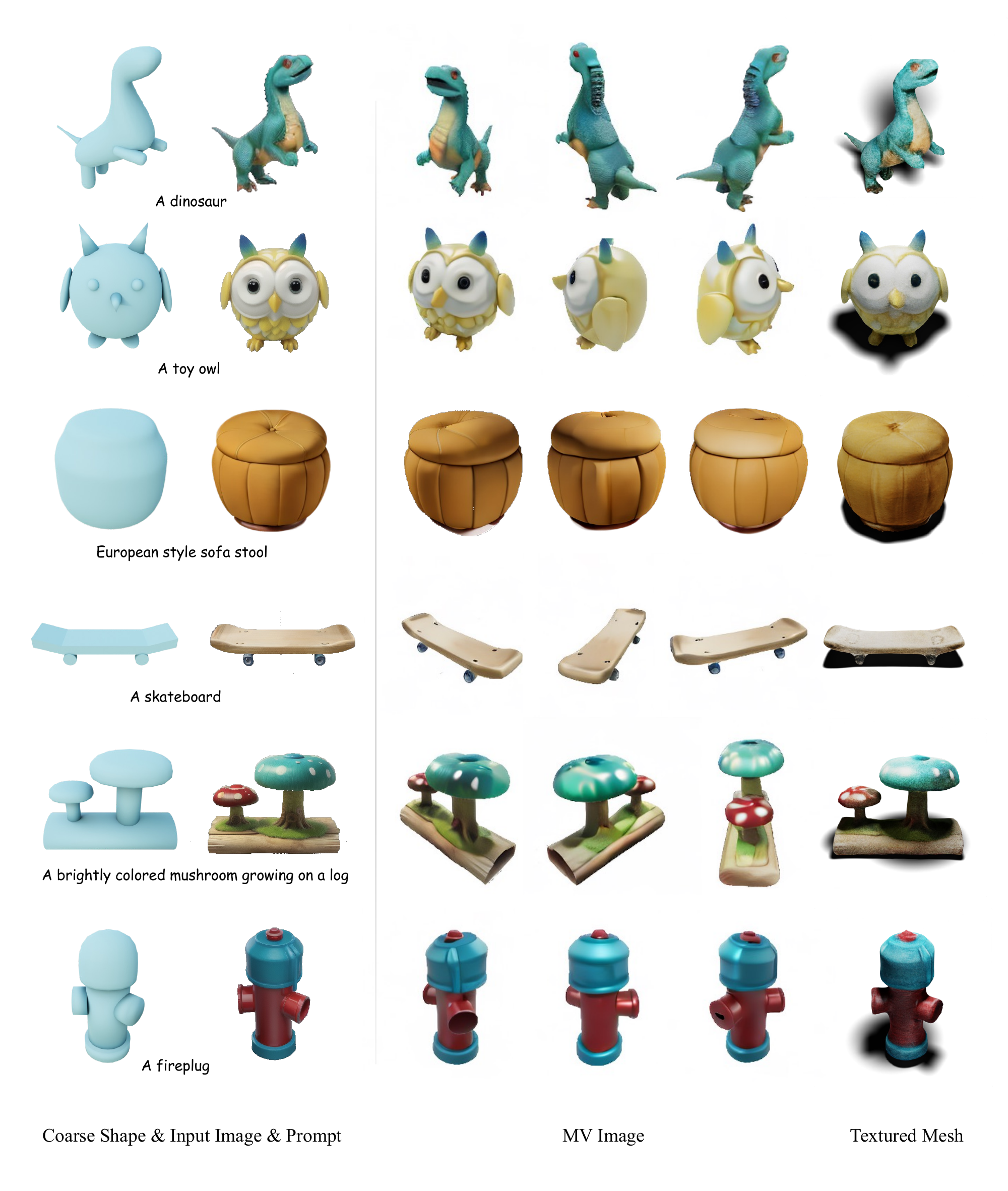}
    \caption{
    More examples of controllable 3D object generation.
    }
    \label{fig:supp_scene_arr_2}
\end{figure*}

\clearpage

\end{document}